\documentclass[aps,twocolumn]{revtex4-1}
\usepackage{amsmath,amssymb}
\usepackage{graphicx}				
\def\mean#1{\left< #1 \right>}
\begin{document}
\title{Cage properties and its implication to the existence of glass transition in hard sphere systems}
\author{Moumita Maiti}
\affiliation{Institut f{\"u}r Theoretische Physik 1, Friedrich-Alexander-Universit{\"a}t Erlangen-N{\"u}rnberg, Staudtstr. 7, 91058 Erlangen, Germany}
\begin{abstract}
In deep supercooled liquids, particles get trapped in transient cages made up of neighbouring particles. Here we define a cage from a geometrical quantity, free volume, such that the free volume of a particle is the cage volume. First we show that the relationship between the average cage volume and the structural relaxation time questions the existence of glass transition in hard sphere systems. Our observation suggests that the cage volume is zero at the transition. Further we show that cage rearrangements are strongly coupled to the single particle squared displacements. Additionally a cage can rearrange by losing its neighbours with almost no change in particle displacements. The picture presented here also supports the complex scenarios of relaxation, dynamic heterogeneity and cooperative rearrangement. 
\end{abstract}
\maketitle
\section{Introduction}
Glass is a fascinating disordered solid, it is mechanically rigid, like a solid, but has a liquid like structure. When crystallization is avoided by a fast quench, the relaxation time of the supercooled liquid increases dramatically by approaching a critical point for many systems \cite{angell,debenedetti,richardetal,hunterandweeks,puseyandmegen}, which is usually called the glass transition point. Hard sphere system is a simple model system to exhibit the glass transition by increasing density, which has been studied extensively in theory \cite{parisiandzamponi}. Experimentally it can be realized using colloidal particles \cite{hunterandweeks,puseyandmegen,weeksetal} or granular materials \cite{richardetal}. However, the dramatic increase of the relaxation time is poorly understood. The relaxation time near the glass transition is too high to measure either in simulations or experiments. There are a few functional forms \cite{goetze,schweizer} proposed to fit the available data on relaxation time as a function of density. The fitting functions are formulated such that from its extrapolation to higher densities relaxation time diverges at the glass transition. This implies that relaxation is impossible at the glass transition point.\\
A series of attempts \cite{cohenandturnbull,turnbullandcohen,cohenandgrest} exist in the literature to understand the transport properties near the glass transition from the free volume theory. The theory gives a relation between diffusivity and the structural variable free volume. Here, we establish a relation between the structural relaxation time and the free volume. We define a cage such that the free volume of a particle is the cage volume. The connection between the average cage volume to the structural relaxation time concludes that the relaxation time is too high when the free volume is zero. This point of zero free volume and very high relaxation time represents here the glass transition point. Then, the relation between the free volume and the density questions the existence of the glass transition in hard sphere systems.\\    
The structural relaxation of a supercooled liquid exhibits particle caging which is absent in the normal liquid. The rearrangement of a cage of a particle is known to happen by hopping of the particle to a new cage is seen by exhibiting a jump in the squared displacement of the particle. The rearrangements of cages have been studied both in the experiments \cite{weeksandweitz1,weeksandweitz} and in the computer simulations \cite{doliwaandheuer,ciamaraetal} where the information of cages are extracted from a dynamical quantity. A cage of a particle here is made up with its neighbours specific to enclose its free volume surface. We show that the cage rearrangements indeed happen when the squared particle displacements exhibit jump. In the same time there are some cases where the rearrangements happen with almost no change in the particle displacements.\\  
A couple of complex scenarios \cite{berthierandbiroli} exist in the relaxation process near the glass transition. At a given time, the particles which have escaped the cage, can be thought as fast moving particles. There are particles at the same time which did not escape the cage and have small displacements. Hence, there is a coexistence of slow and fast moving particles in space \cite{weeksetal,glotzer,ediger,cammarotaetal}, known as dynamic heterogeneity. The fast moving particles cluster in space, specifically form a string-like cluster \cite{kobetal,donatietal1,donatietal}. A recent work \cite{starretal} has claimed that the string like clusters are the cooperative rearranging region where more than one particle rearranges cooperatively. We ensure here that our picture of cage rearrangements also supports these complex scenarios of the relaxation process.\\ 
\section{Simulation}
The system consists of hard spheres binary mixture, there are $\frac{N}{2}$ numbers of particles with diameter $\sigma$ and rest $\frac{N}{2}$ numbers of particles with diameter $1.4\sigma$. The binary mixture is considered to avoid crystallization.  The starting configuration is prepared by performing the conjugate gradient minimization of randomly initialized configuration with the repulsive harmonic potential $V(r) = (1 - \frac{r_{ij}}{\sigma_{ij}})^2$ for $r_{ij} < \sigma_{ij}$, and zero otherwise. After the minimization the configuration is the overlap free hard sphere configuration. Then using this as starting configuration We conduct event driven molecular dynamics simulation \cite{rapaport} to equilibrate the system at different packing fractions $\phi = \frac{N\pi(\sigma_{1}^3 + \sigma_2^3)}{12L^3}$. $\sigma_1 = \sigma$, $\sigma_2 = 1.4\sigma$ and $L$ is the length of a cubic box. Several packing fractions $\phi = 0.52, 0.53, 0.54, 0.55, 0.56, 0.57, 0.58, 0.59$, and $0.595$ are considered. The system size is $N = 2000$. The runs are long enough to produce the mean squared particle displacement $2\sigma$ for the highest density $\phi = 0.595$, and more than $100\sigma$ for lower densities. All the results presented here are in the reduced units, {\it i.e.} length in the unit of $\sigma$ and time in the unit of $\sigma/(k_{B}T/m)^{1/2}$.
\section{Results}
\begin{figure}[htbp] 
   \centering
   \includegraphics[width=2in]{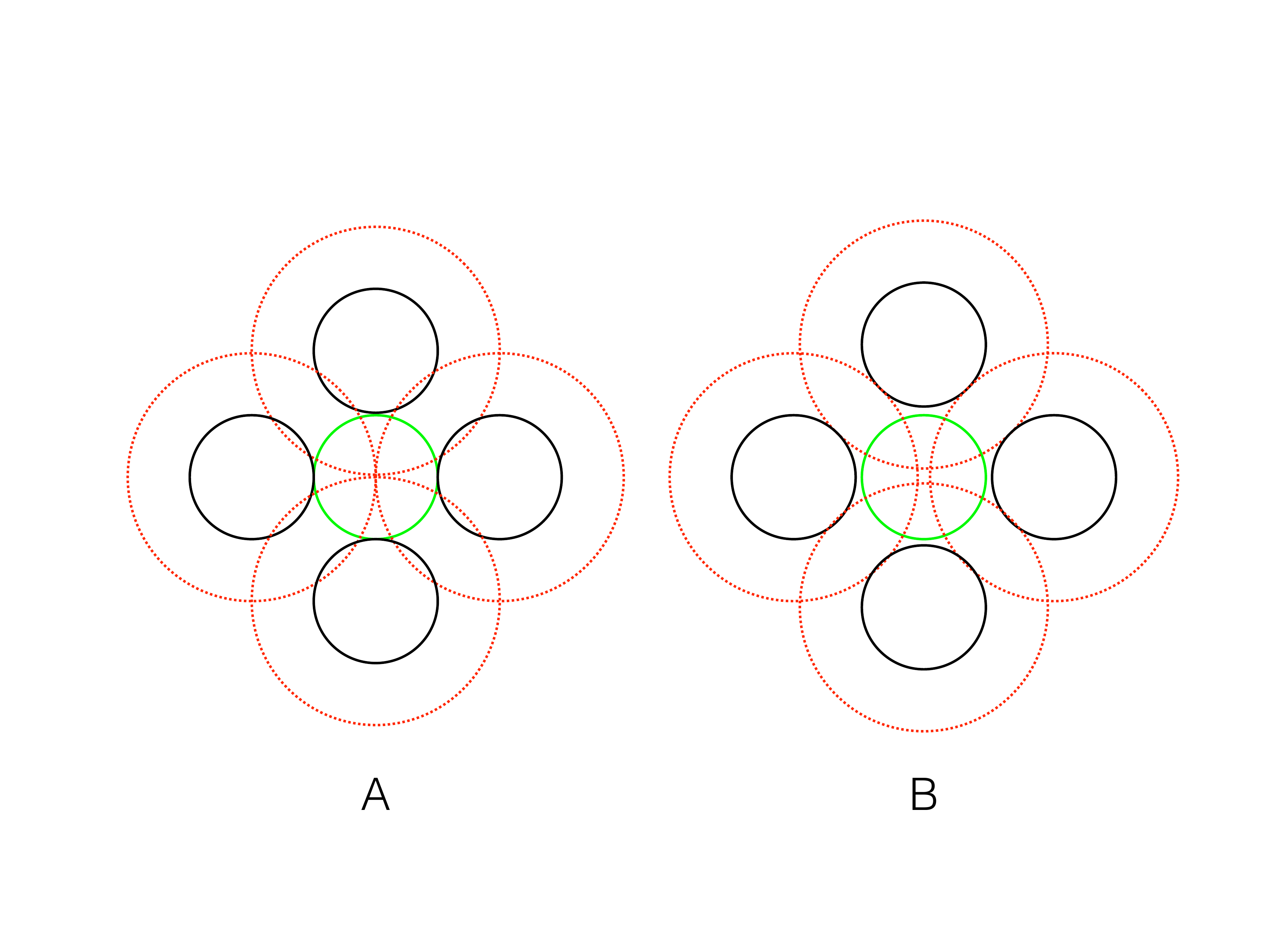}
   \caption{(A) Neighbours of a green particle are in contact, and the free space at its centre is zero. (B) Neighbours of the green particle are away and its centre has finite free space.}
   \label{fig:snap}
\end{figure}

We define a cage surrounded by neighbours of the free volume of a particle. The free volume is an available space to a hard sphere keeping its neighbours positions fixed, which is shown schematically in Fig. ~\ref{fig:snap}. An easy example is that a particle with $2d$ ($d$ dimension) contacts has zero free volume, shown in Fig.~ \ref{fig:snap}(A). The properties of free volume has been explored for equilibrium liquids extensively \cite{sastryetal1,sastryetal2,maitietal}.  A previous work has shown that the free volume of a hard sphere is mostly aspherical \cite{maitiandsastry}, hence the cage defined here is not necessarily spherical. Note here the cage size is not dependending on a cutoff. The cage volume is the free volume of a particle. The relaxation mechanism here will focus on the rearrangement of neighbours of a cage.
\subsection{Relation between the relaxation time and the cage volume}
\begin{figure}[htbp] 
   \centering
   \includegraphics[width=2in]{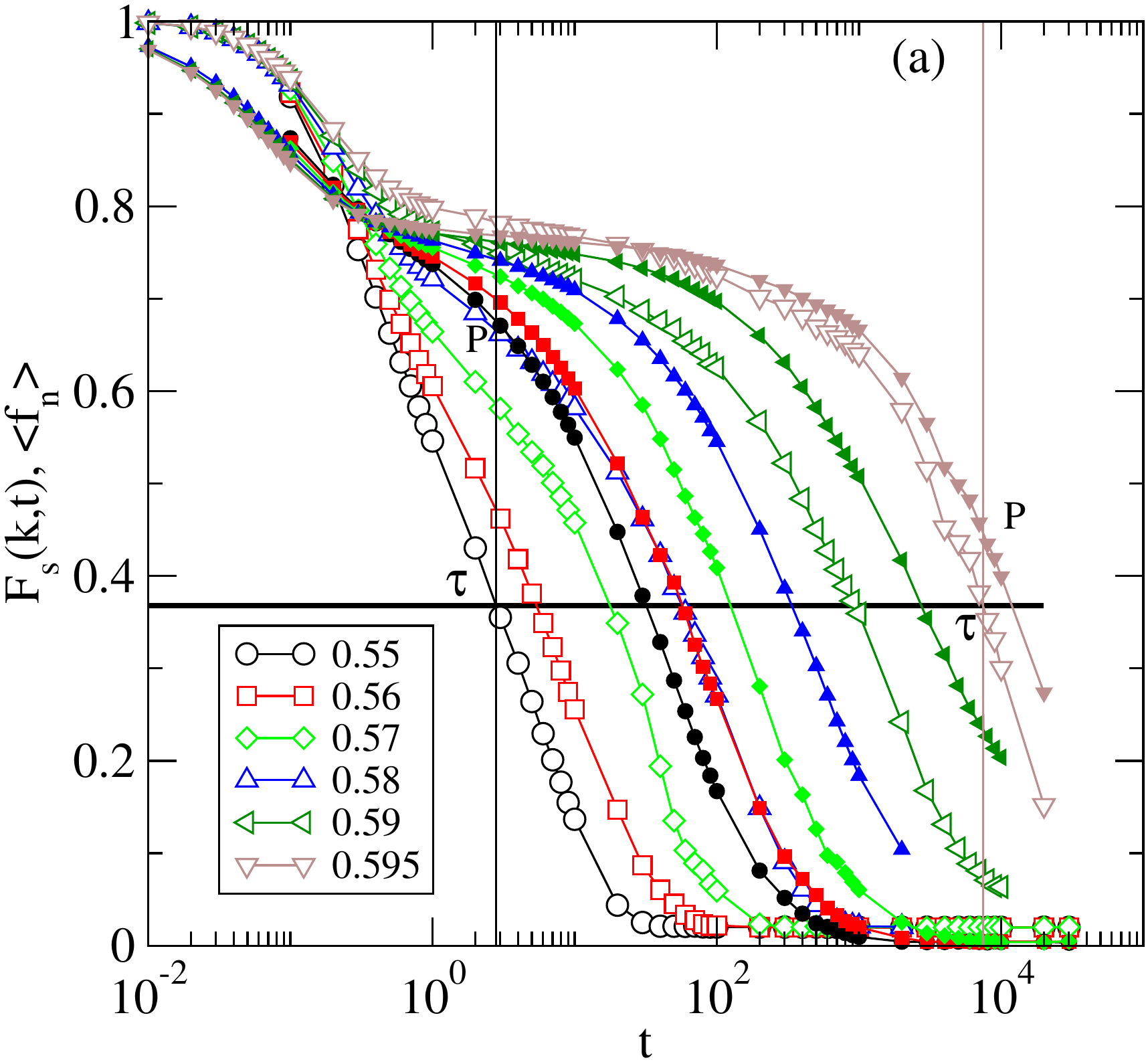} 
   \includegraphics[width=2in]{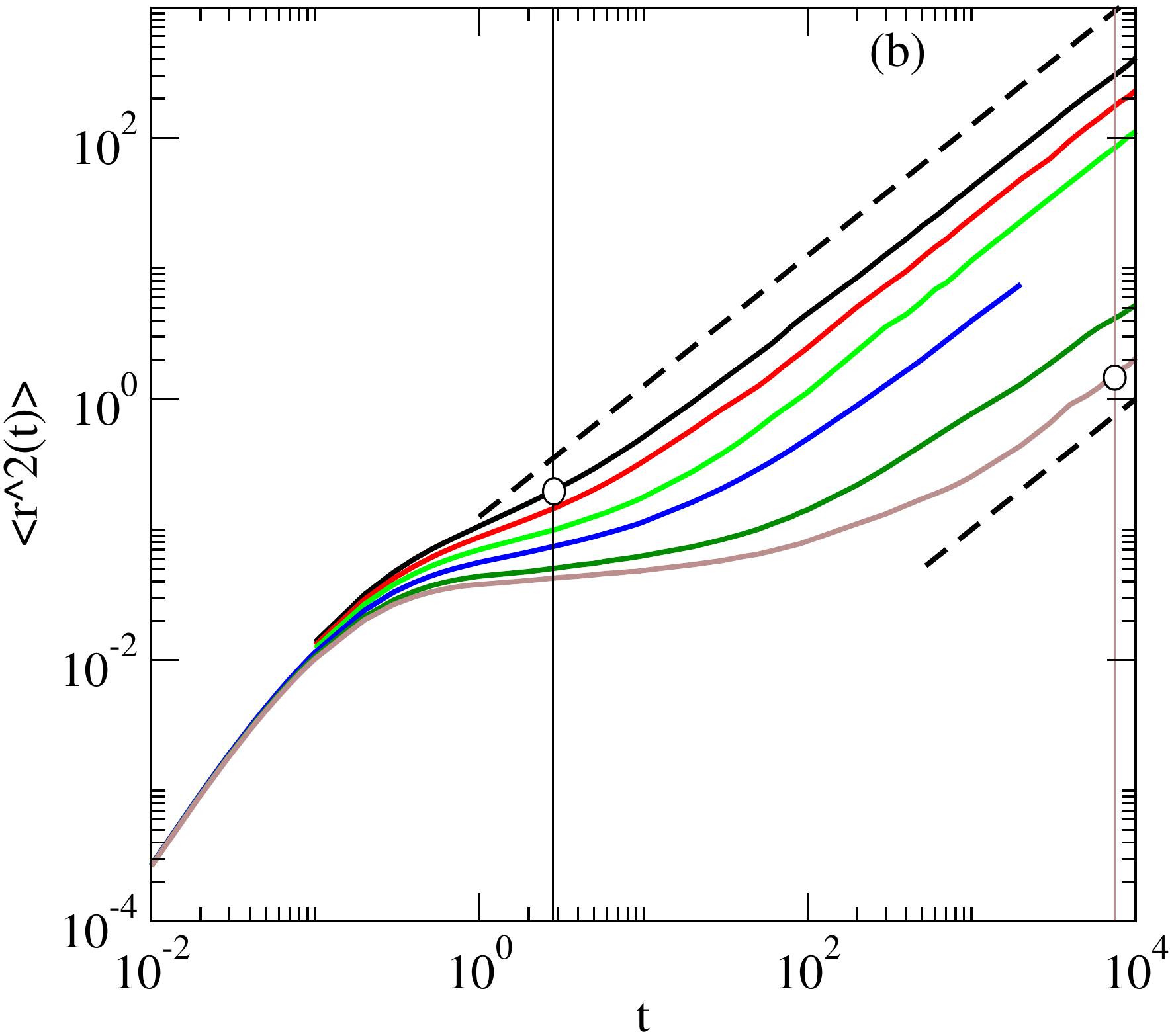} 
   \caption{(a) Self-intermediate scattering function $F_{s}(k,t)$ vs $t$ data are shown by open symbols, and the average of the fraction $f_{n} = \frac{N(t)}{N_{0}}$ are shown by filled symbols for several densities. The relaxation time $\tau$ has the value of $F_{s}(k,t) = 1/e$, a black line at value $1/e$ is drawn to guide our eye. The line at value $t = \tau$ for a given density intersects the plot of $\mean{f_{n}}$ vs $t$ at point $P$. (b) The average of the mean square displacement $\mean{r^2(t)}$ vs. $t$ for several densities, the colouring is same as panel(a). The value of the mean square displacement at the relaxation time $\tau$ is shown by open circle for two densities.}
   \label{fig:fig1}
\end{figure}

First we calculate the self part of the intermediate scattering function $F_{s}(k,t) = \frac{1}{N}<\sum_{i}e^{(i\vec{k}. (\vec{r_{i}}(t) - \vec{r_{i}}(0)))}>$ for the value of $k = 6.1$ where the first diffraction peak of hard spheres is present \cite{Brambillaetal}. $F_{s}(k,t)$ as a function of time $t$ is plotted in Fig.~ \ref{fig:fig1}(a). At low densities the behaviour is exponential. There are three distinct regimes at higher densities, early time the ballistic motion, then there is a plateau reflecting the caged motion of the particle. The late time, the particle escapes the cage and enters into the $\alpha$-relaxation regime which has the stretched exponential behaviour as expected. Now, at higher densities, the mean squared displacement as a function of time $t$ in Fig.~ \ref{fig:fig1}(b) also has three distinct regimes, once the particle escapes the cage it diffuses. The diffusive regime behaves as $\mean{r^2(t)} = 6Dt$, where $D$ is a diffusion constant. In log-log scale the diffusive regime is parallel to the line $\sim t$ shown as a dashed line in the figure. We have shown the data of six packing fractions starting from $\phi = 0.54$.

\begin{figure}[htbp] 
   \centering
   \includegraphics[width=2.5in]{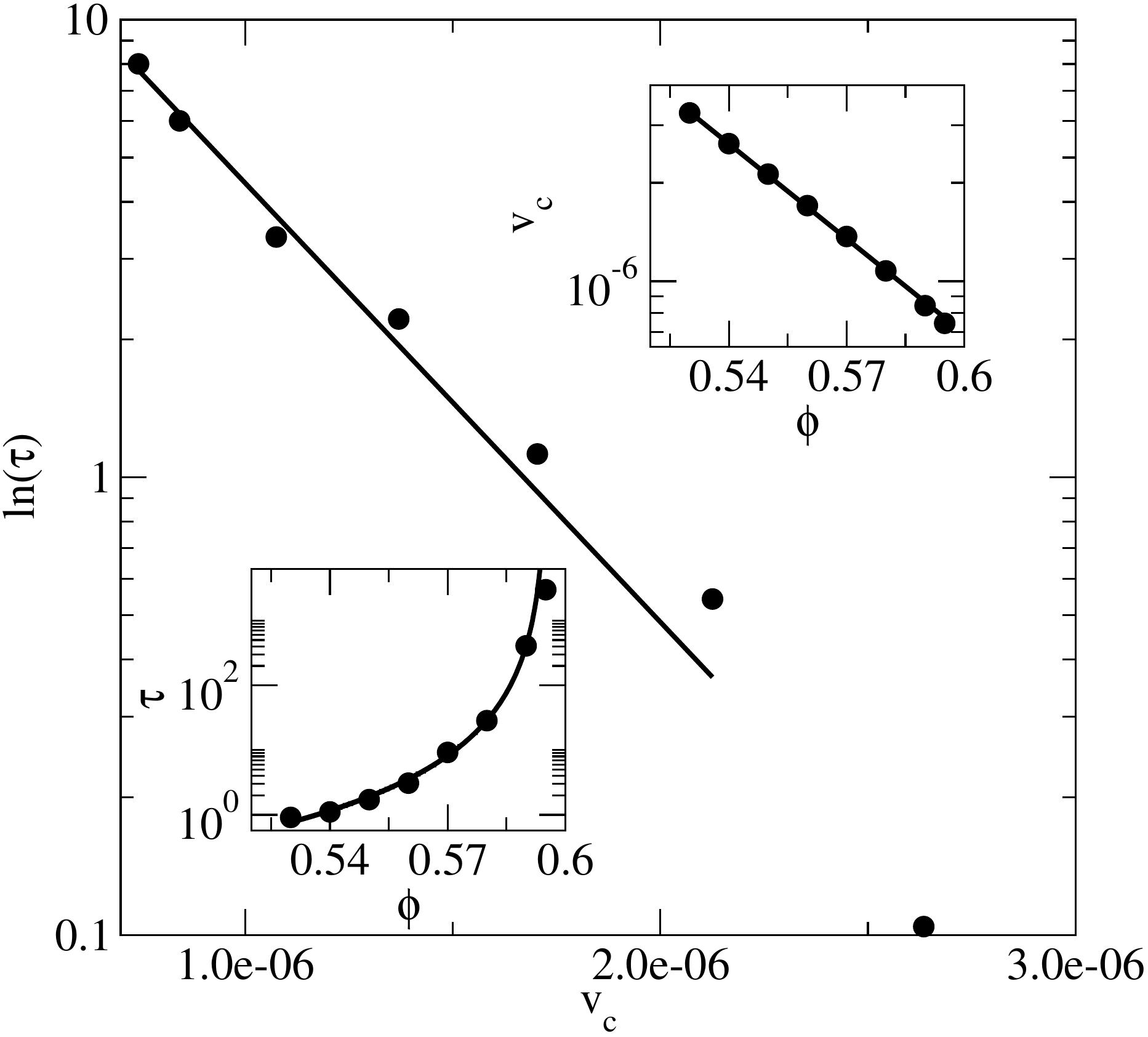} 
   \caption{The logarithm of the relaxation time $\tau$ is plotted against the cage volume $v_{c}$, which is fit with a function $\ln(\tau) = a_{g}\exp(-a1*(v_{c} - v_{c}^{g}))$ for smaller values of $v_{c}$. The one of the insets shows that $v_{c}$ is smaller at higher packing fraction $\phi$ and they have exponential relation. Another inset shows that the relaxation time $\tau$ as a function of packing fraction $\phi$. The power law fit of the mode coupling prediction gives the critical density $\phi_{c} = 0.5953$.}
   \label{fig:fig2}
\end{figure}

We calculate the $\alpha$-relaxation time $\tau$ by setting $F_{s}(k,\tau) = \frac{1}{e}$, which is shown in Fig. ~\ref{fig:fig1}(a) by a black line running parallel to the time axis. We observe that $\tau$ increases rapidly by increasing the packing fraction shown in Fig. ~\ref{fig:fig2} inset. The fitting of the relaxation time data to the mode coupling prediction \cite{goetze} $\tau \sim (\phi_{c} - \phi)^{\gamma}$ gives the density $\phi_{c} = 0.5953$ with $\gamma = -2.5$. There are other possible functions to fit the data like the Vogel-Fulcher-Tammann(VFT) \cite{schweizer} form and will give a different estimation for the singular point $\phi_{c}$. We stick here to one functional form as it is not our goal to find out the true glass transition density. The mode-coupling functional fit claims that there is a singularity at $\phi_{c} = 0.5953$ even though the relaxation is possible beyond $\phi_{c}$ \cite{Brambillaetal}.

Now, the plateau in the mean squared displacement represents the cage size, here we have a better quantity the cage volume as the cage here is well defined. The relaxation time $\tau$ and the average cage volume $v_{c}$ are calculated for a given density and plotted against each other for densities $\phi = 0.54-0.595$ in Fig.~ \ref{fig:fig2}. We find that the relaxation time and $v_{c}$ have a well defined relation $\ln(\tau) = a_{g}\exp(-a1*(v_{c} - v_{c}^{g}))$, such that $\tau = \exp(a_g) = 10^{17}$ at $v_{c} = v_{c}^{g} = 10^{-12}$. If we look at the variation of the average cage volume $v_{c}$ as a function of packing fraction $\phi$ in the inset of Fig.~ \ref{fig:fig2}, we find that it has value around $10^{-6}$. Hence, $v_{c}^{g} = 10^{-12}$ can be considered as zero. If we assume at the glass transition density $\phi = \phi_{g}$, the relaxation time is as high as $10^{17}$, such that the relaxation by simulation is not accessible. Then at $\phi_{g}$ average free volume $v_{c}^{g}$ is zero. The exponential relation of $v_c$ with $\phi$ can not estimate any meaningful glass transition density $\phi_{g}$ where $v_{c} = 0$. It questions the existence of the glass transition for hard spheres. Now the power law distribution of the free volume at jamming \cite{maitiandsastry} gives the value of average free volume finite, hence also the jamming point can not coincide with the glass transition density. Note, for molecular glass it was shown that the free volume is zero at the VFT critical temperature \cite{starretal1}. Our results imply that the single particle quantity free volume or the cage volume can give us the relaxation time upto the density $\phi = 0.595$, one can describe the relaxation behaviour in the single particle level. 

\subsection{Cage rearrangements in the structural relaxation}
Motivated by our finding, we now define a new single particle quantity $f_n = \frac{N(t)}{N_0}$, where $N_0$ is the number of neighbours of the cage at $t=0$. We tag the neighbours at $t=0$ and find out $N(t)$ which is the number of survived neighbours out of tagged neighbours $N_0$ at a given time $t$. The variation of $f_n$ with time describes the rearrangement of cage neighbours around a particle. We plot $\mean{f_n}$ in Fig. ~\ref{fig:fig1}(a) as a function of time in the same panel as the self-intermediate scattering function.  The quantity $\mean{f_{n}}$ also exhibits three distinct regimes at higher densities similar as $F_{s}(k,t)$, even the late time behaviour can be fit to the stretched exponential as seen clearly in the figure by being parallel to the corresponding $F_{s}(k,t)$ $\alpha$-relaxation regime. Fig. ~\ref{fig:fig1}(a) shows that the plateau amplitude of the fraction $\mean{f_{n}}$ increases with increasing the density. So, longer the caging time, higher is $\mean{f_n}$ which indicates the cage surrounded by higher fraction of initial neighbours take longer time to escape the cage. The higher value of $\mean{f_{n}}$ at plateau by increasing the density means that the cage there is more tightly packed when a particle escapes from it. Intuitively the probability of relaxation by hopping process is high if the cage during the process of escaping is tightly packed. This implies that mostly particles relax by escaping the cages following jumps at higher densities.\\
\begin{figure}[htbp] 
   \centering
   \includegraphics[width=2.25in]{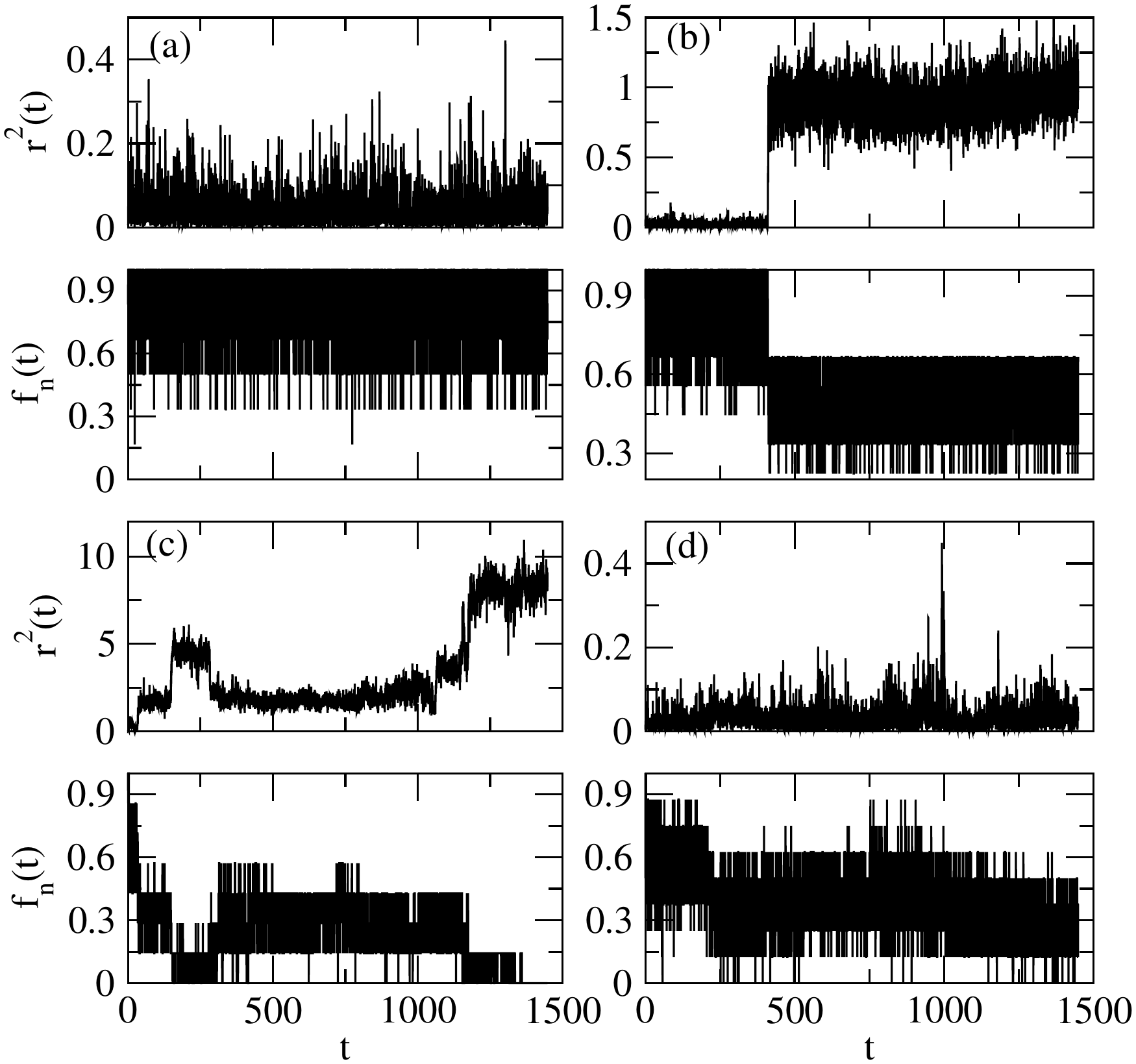} 
   \caption{Squared displacements of individual particles are shown in four panels (a), (b), (c), and (d). The corresponding data of $f_{n}$ is shown below each panel. We observe $f_{n}$ behaves according to the displacement, {\it i.e.} when it drops suddenly, it is due to the jump in the squared displacement. Again the backward jump increases the value of $f_n$ as seen in the panel(c). Only panel (d) shows a particle which has no change in the displacement, but there is a drop from the average value $\sim 0.6$ to $\sim 0.3$ in $f_{n}$.}
   \label{fig:fig3}
\end{figure}
A line at $t = \tau$ parallel to the $y$-axis is drawn for a given density which intersects the plot of $\mean{f_{n}}$ at a point $P$, shown in Fig. ~\ref{fig:fig1}(a). The $y$-value of $P$ will tell us how many neighbours of time $t=0$ a particle looses in order to relax. Surprisingly, the value $P_{\tau}$ {\it i.e.} $\mean{\frac{N(\tau)}{N_0}}$ is not constant with the packing fraction $\phi$, it decreases with increasing packing fraction. Higher the density, higher is the $\mean{f_n}$ at the plateau but interestingly lower is $\mean{f_n(\tau)}$. If the relaxation takes place by hopping, then the fraction $f_n$ after relaxation will be smaller than any other way of relaxations. Having lower $\mean{f_n(\tau)}$ at higher densities supports our anticipation that particles at higher densities relax mostly by hopping process. As the jump in particle squared displacement is interpreted as hopping, clearly at higher densities the displacement of the particle should be tightly correlated with the particle cage rearrangement $f_n$. \\
\subsection{The connection between the cage rearrangements and squared particle displacements}  
To investigate further the squared displacements of individual particles are considered. Fig. ~\ref{fig:fig3} shows four cases of squared particle displacements  during the time of observation: (a) a particle is caged, (b) a particle escapes the cage following a single jump, (c) a particle escapes the cage following successive jumps, and (d) another case of a caged particle. For case(a) we observe that there is no neighbour rearrangement in the time of observation. The case(b) has neighbour rearrangement by dropping $f_n$ to an average value $0.4$ exactly at same time where the jump is observed. Case(c) shows how strong is the correlation between $f_n$ and the squared displacement $r^2$, $f_n$ drops to an average value $0.3$ at a point where the first jump is present. After the second jump $f_{n} = 0$ almost. Afterwards the particle has a backward jump and at last another forward jump, the corresponding $f_{n}$ behaves consistently with this. For case(d) interestingly the particle did not displace much to escape the cage, but there is neighbour rearrangement. Then the question arises whether the squared displacement $r^2$ and $f_n$ is always correlated. \\
\begin{figure}[htbp] 
   \centering
   \includegraphics[width=1.5in]{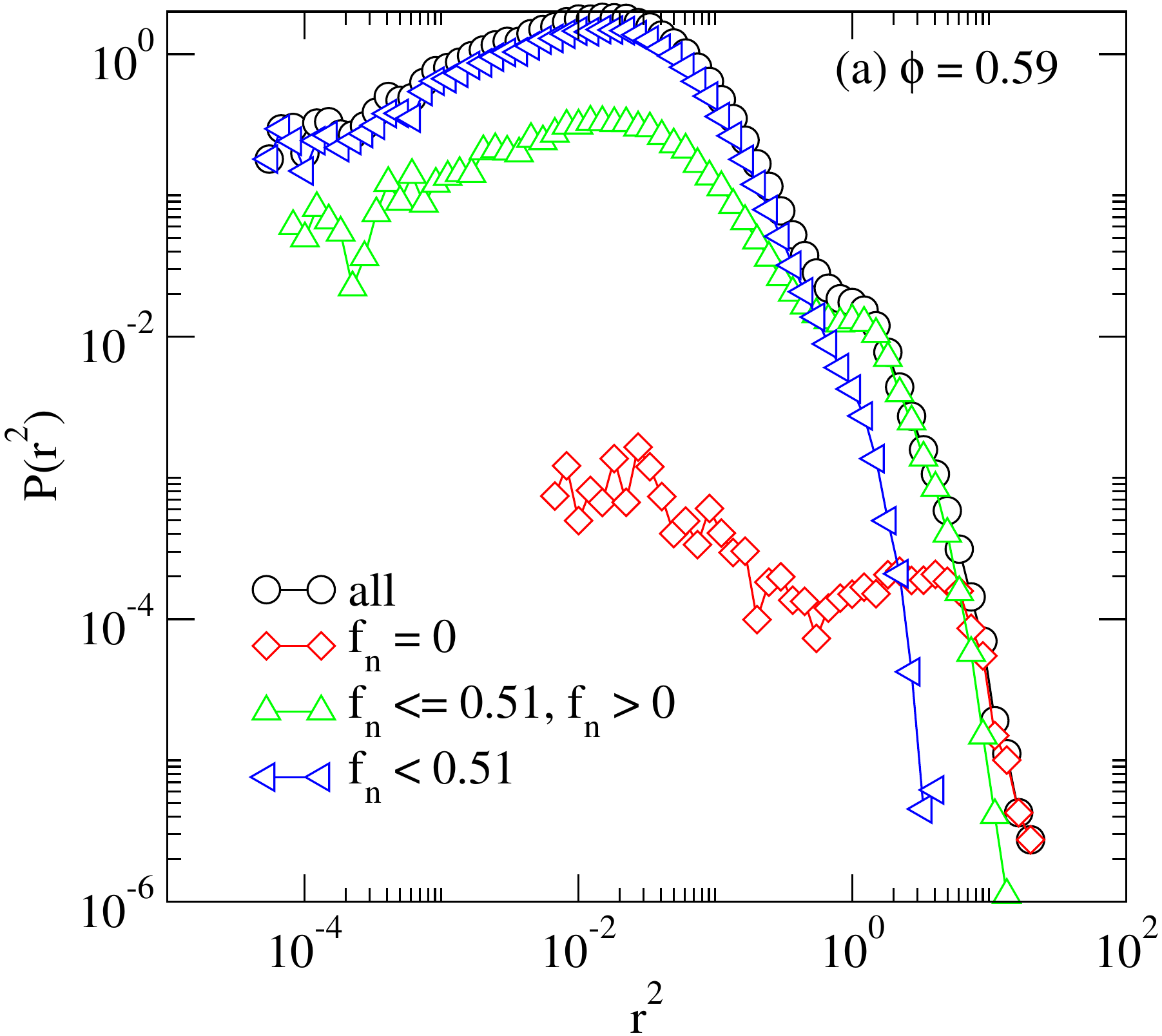}
    \includegraphics[width=1.5in]{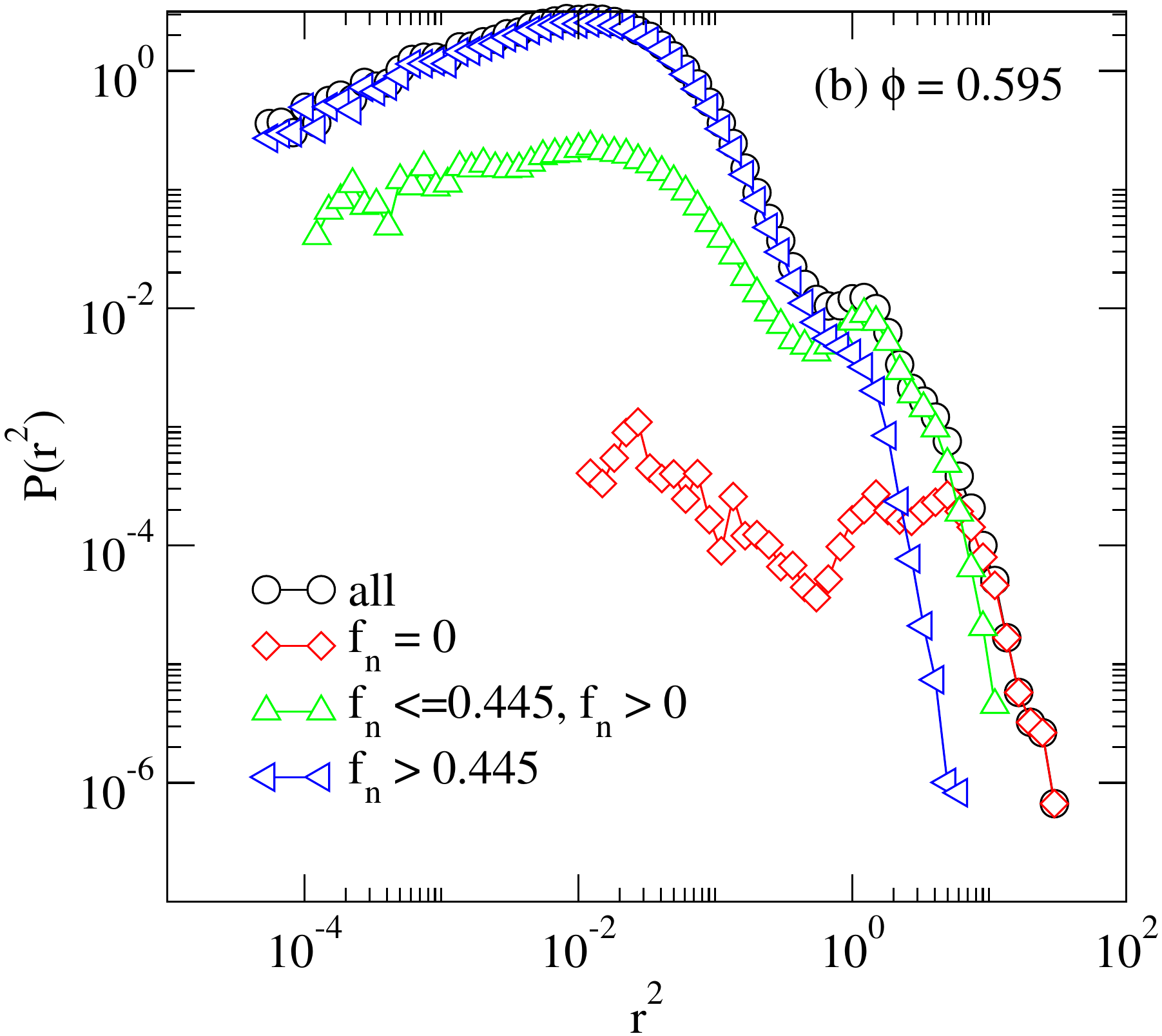}
    \includegraphics[width=1.5in]{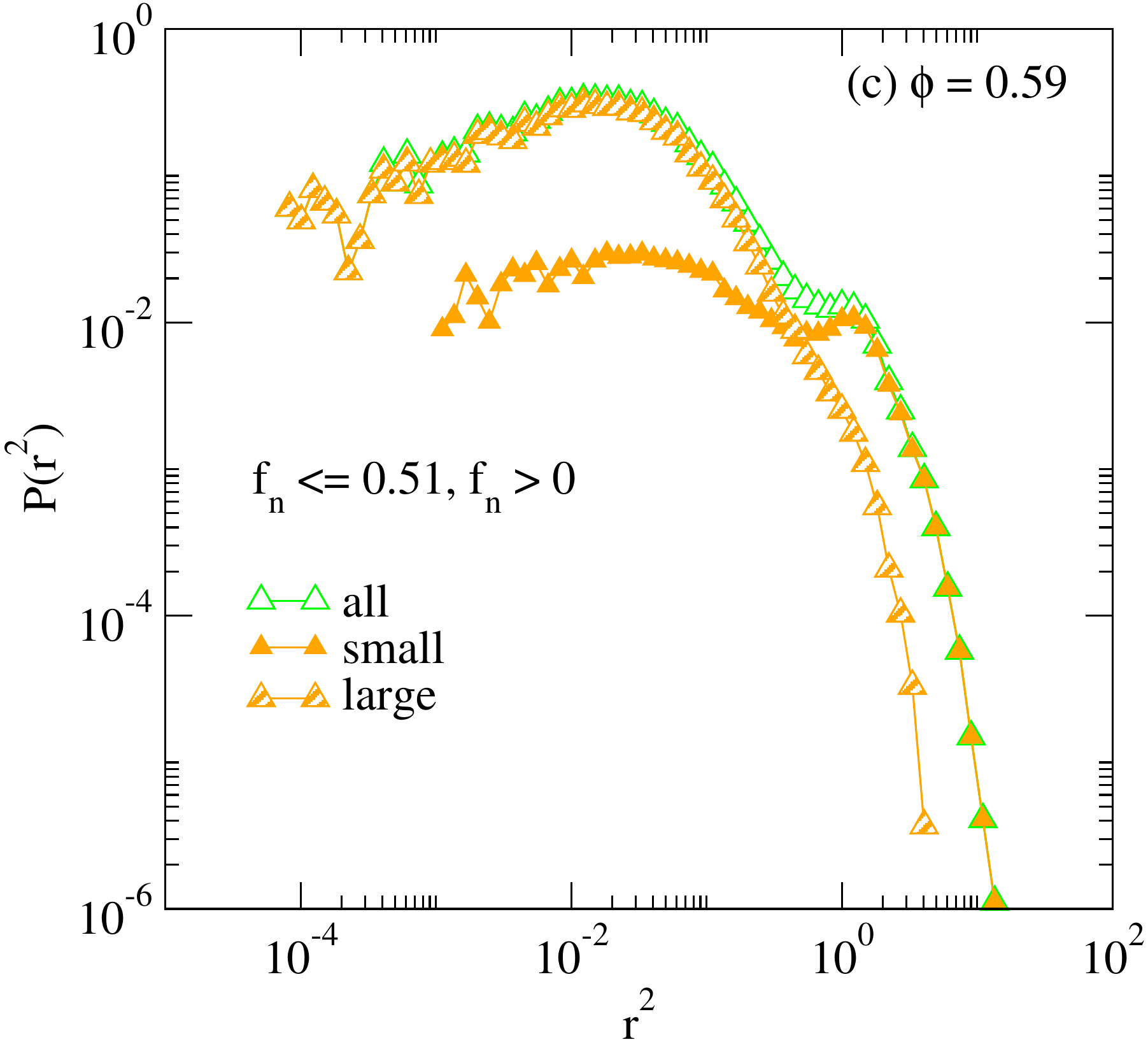}
    \includegraphics[width=1.5in]{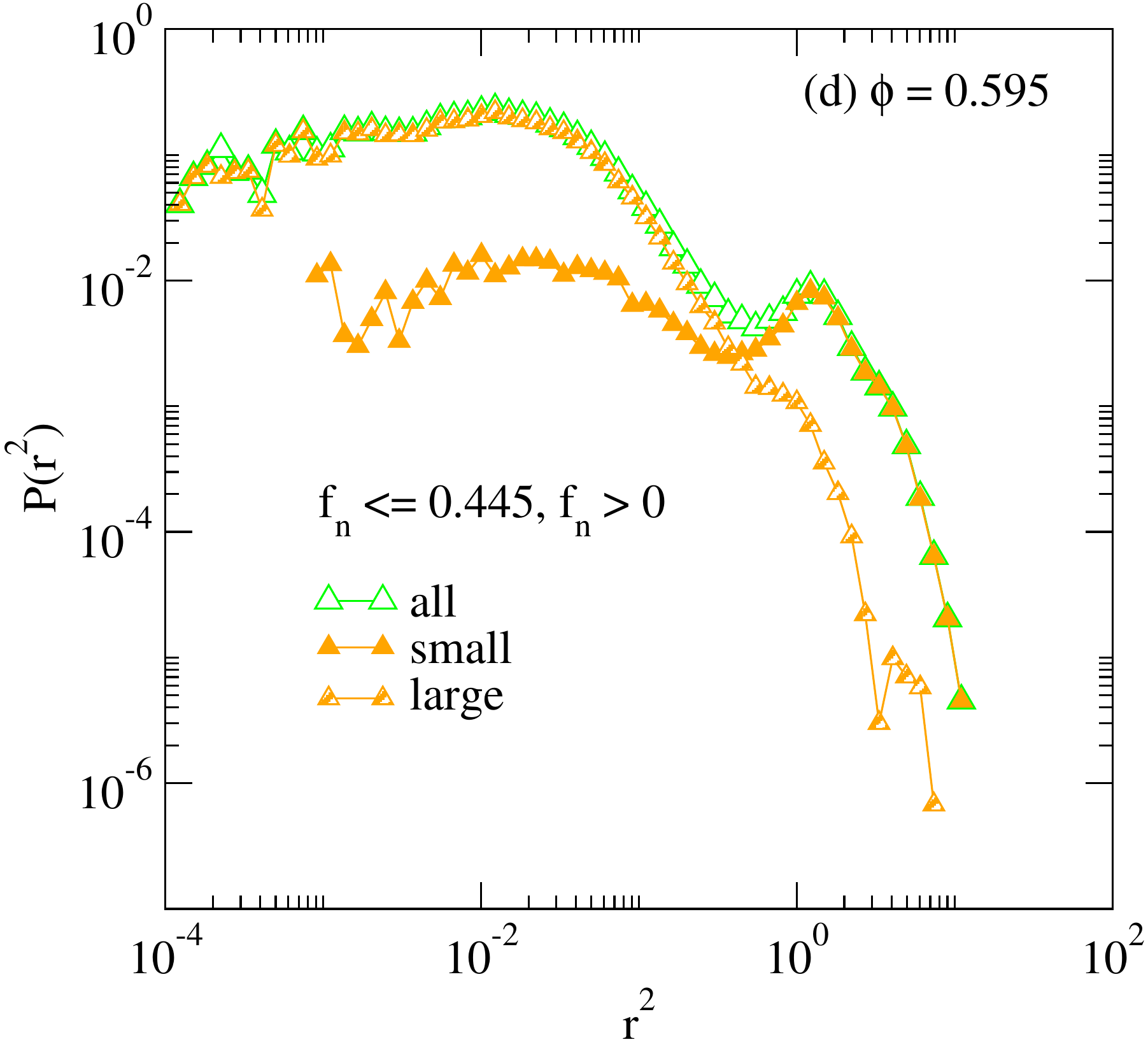}
   \caption{The distribution of the particle squared displacement $r^2$ as a function of $r^2$ at densities (a) $\phi = 0.59$ and (b) $\phi = 0.595$ at time $t_{\alpha}$(see text for description). We define three regimes of $f_{n}$, $f_{n} = 0$, $f_{n} \le \mean{f_{n}(\tau)}, f_{n}>0$, and $f_{n} > \mean{f_{n}(\tau)}$. Then we plot the displacement of particles falling in these three regimes. We consider the particles falling in the regime $f_{n} \le \mean{f_{n}(\tau)}, f_{n} > 0$ separately in panels (c) for the density $\phi = 0.59$ and (d) $\phi = 0.595$. Here we show that the large particles have mainly the slow displacements even though with the fraction $\mean{f_{n}(\tau)}$.}
   \label{fig:fig4}
\end{figure}
In order to investigate that we choose two densities $\phi = 0.59, 0.595$. The distribution of squared displacement is calculated at a given time $t_{\alpha}$ where the system escapes the caging regime and enters into the $\alpha$-relaxation regime. The distribution of $r^2(t_{\alpha})$ shows in Fig. ~\ref{fig:fig4}(a,b) that there are more than one peak which has been observed before as well \cite{kumaretal}. Fig. ~\ref{fig:fig1}(a) gives the value of $\mean{f_{n}(\tau)} = 0.51$ and $\mean{f_{n}(\tau)} = 0.445$ for densities $\phi = 0.59$ and $0.595$. We name particles `R' if they have the value of $f_{n} < \mean{f_{n}(\tau)}$. Now, the fraction $f_n = \frac{N(t)}{N_0}$ is binned into three regimes: (a) $f_n = 0$, (b) $f_n< \mean{f_{n}(\tau)}$ , and (c) rest. We have observed in Fig. ~\ref{fig:fig3}(c) that successive jumps can give $f_n = 0$, and should have very high squared displacement. Now particles sitting in three regimes are picked up and plot the corresponding distribution $r^2(t_{\alpha})$ in the same panel Fig. ~\ref{fig:fig4}(a,b). Particles with $f_n = 0$ are indeed those which have higher displacements so they are multiple times `hoppers'. `Hoppers' are those which have the squared displacement larger than the value of $r^2(t)$ of first minima, $r^2_{min}(t)$.\\
Firstly the particles falling into the regime of (b) contributes to the peak at large displacement, and of regime (c) has only one peak at small displacement. This observation confirms that the particles which exhibit jumps(either one or multiple) in the single particle squared displacement, also exhibits at same time sudden drop in $f_n$, {\it i.e.} the neighbours rearrangement of particles is tightly bound to the squared particle displacements. So, along with dynamical heterogeneity there is spatial heterogeneity of local structural rearrangement. Moreover, the number of rearranged particles decreases with increasing density, but the number of `hoppers' exhibiting multiple jumps$(f_n = 0)$ increases by increasing density. This means that `hoppers' with successive jumps are the ones to relax the system at further densities. Interestingly the case of (b) where particles have relaxed or rearranged, also have a displacement peak even below $r^2_{min}(t)$, is supporting the scenario of Fig. ~\ref{fig:fig3}(d). One possible explanation is that there are some `slow' particles surrounded by `hoppers'. This result indicates that not necessarily a particle has to move out of the cage in order to relax. If there are a few `hoppers', they assist as well their neighbours to rearrange, so there are particles more than number of `hoppers' which have relaxed in time $t$. At time $t = \tau$, some of all particles with $f_n < \mean{f_n(\tau)}$ are `hoppers' and some are `slow' particles. So, the possibility of being dynamical heterogeneous at $t = \tau$ can not be ignored specially at those higher densities where $\mean{f_{n}}$ is small. Next Fig. ~\ref{fig:fig4}(c,d) shows that the `hoppers' are mainly particles small in size, and the big particles rearrange due to small neighbours around them. So, the polydispersity plays a role to the dynamic heterogeneity.\\
\begin{figure}[htbp] 
   \centering
   \includegraphics[width=2in]{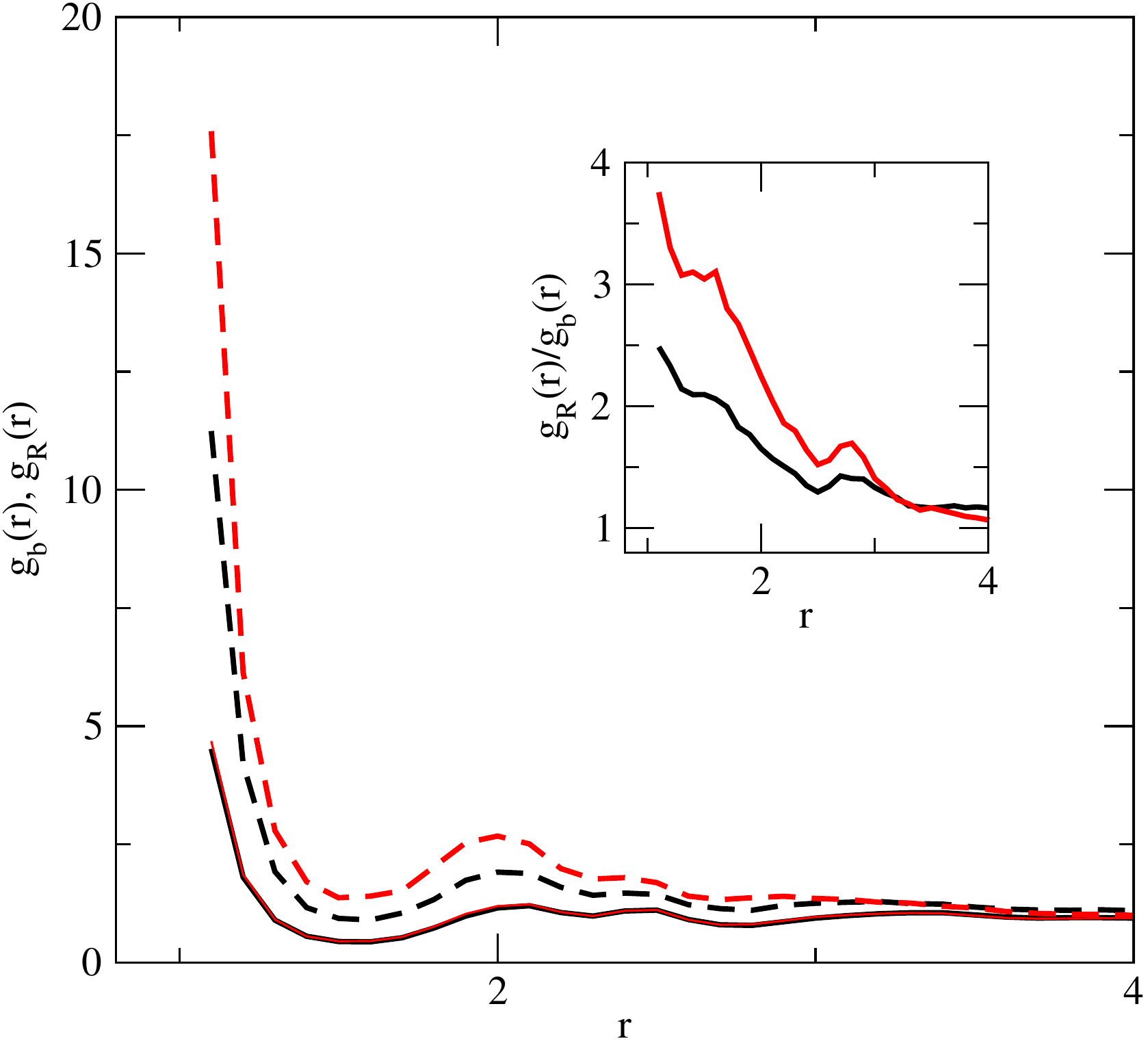} 
   \caption{Radial distribution function of bulk particles $g_{b}(r)$(solid line) and relaxed particles $g_{R}(r)$(dashed line) of packing fractions $\phi = 0.59$(black) and $0.595$(red). The inset shows that the ratio between $g_{R}(r)$ and $g_{b}(r)$.}
   \label{fig:fig5}
\end{figure}
\subsection{Clustering of rearranged particles}
It has been shown that fast and slow moving particles cluster in space \cite{kobetal,donatietal1,donatietal}, the work in \cite{kobetal} shows that fast moving particles have strong radial distribution peaks compared to the radial distribtution function of bulk particles. We pick up the relaxed particles `R' and calculate the radial distribution function $g_{R}(r)$. We show in Fig. ~\ref{fig:fig5} that the radial distribution function of bulk particles $g_{b}(r)$ for two densities are same. But $g_{R}(r)$ is different and the peaks are stronger at higher densities.  The ratio $g_{R}(r)/g_{b}(r)$ increases with increasing density shown in Fig. ~\ref{fig:fig5} inset. The cooperative rearrangement is stronger at higher densities. One can define the cluster of `R' particles as cooperatively rearranging region `CRR'. So, the single particle picture presented here can explain the cooperativity in the structural relaxation.
\begin{figure}[htbp] 
   \centering
   \includegraphics[width=2in]{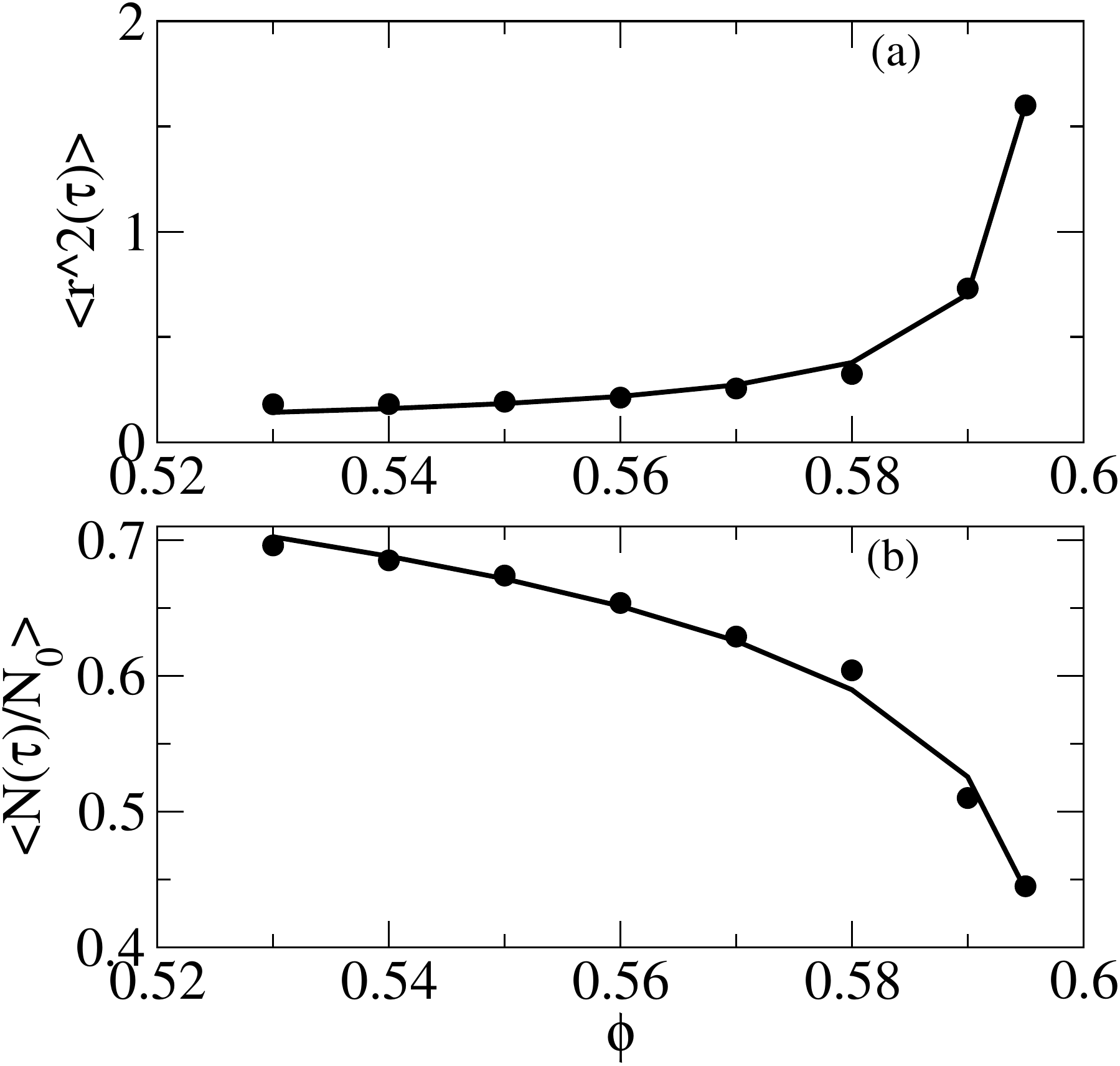} 
   \caption{(a) The mean squared displacement at the relaxation time, $\mean{r^2(\tau)}$ as a function of the packing fraction $\phi$. The data is fit with the function $\sim (\phi_{c} - \phi)^{\alpha}$ with $\alpha = -0.72$ and $\phi_{c} = 0.598$. (b) The average $\mean{f_{n}(\tau)}$ as a function of packing fraction $\phi$. Here the fit function is $\sim (\phi_{c} - \phi)^{\nu}$ and $\nu = 0.13$ and $\phi_{c} = 0.5967$.}
   \label{fig:fig6}
\end{figure}
\section{Discussion}
For hard sphere systems there is a well defined relation between the average free volume and the structural relaxation time. This relation yields very high relaxation time where the average free volume comes out to be zero. We call this point the glass transition density, then the relation between the average free volume and the density doubts the existence of the glass transition in hard sphere systems.\\

We have shown that the single particle variable $f_n$ which is derived from a structural quantity, the neighbours of a free volume, supports the hopping process in the structural relaxation at higher densities. There are particles which have small displacements, still their cages may rearrange. A possible scenario is that these particles are surrounded by `hoppers'. So, not necessarily a cage rearranges if the particle is fast moving. Hence, even complete cage rearrangements can keep in tact the dynamic heterogeneity at $t = \tau$. If we pick up particles whose cages have rearranged in a duration $t$, they cluster, this implies the cage rearrangements support as well the cooperative rearrangements in the structural relaxation.\\

Lastly, from our simulation data we speculate a relaxation mechanism at the mode coupling critical density. We assume that at the critical density $\phi = \phi_{c}$, $\mean{\frac{N(\tau)}{N_0}} = 0$, then the fitting the data(Fig. ~\ref{fig:fig6}(b)) with $\sim (\phi_{g} - \phi)^{\nu}$ inspired from the mode-coupling prediction gives $\phi_{c} = 0.5967$ and $\nu = 0.13$. The value of $\phi_{c}$ is close to the packing fraction obtained from the fitting of the relaxation time $\tau$ as a function of packing fraction $\phi$, which supports our assumption. As discussed before $\mean{\frac{N(\tau)}{N_0}} = 0$ represents multiple jumps in the squared particle displacement, then at $\phi_{c}$ mostly `hoppers' have multiple jumps.\\

Now, we mark the mean squared displacement $\mean{r^2(\tau)}$ in Fig. ~\ref{fig:fig1}(b). Surprisingly, this quantity increases with increasing density shown in Fig. ~\ref{fig:fig6}(a) and we show the data can be fit with  the function $\sim (\phi_c - \phi)^{\alpha}$  where $\phi_{c} = 0.598$ and $\alpha = -0.72$, again almost the same density as obtained from the fit to the relaxation time. This implies that $\mean{r^2(\tau)}$ diverges at $\phi = \phi_{c}$. The possible explanation is that the `hoppers' exhibiting successive jumps are strongly correlated in space, such that the cluster of `hoppers' is most probably `string' like and it has an infinite length. The same argument has been given before \cite{donatietal} by calculating the string length formed by fast moving particles as a function of temperature for the Lennard-Jones system.\\

In conclusion a single particle picture presented here can explain most of the existing findings of the relaxation process upto the mode coupling critical temperature. The picture with more than one particle may be required to describe the system above the density $\phi = 0.595$ and can be extended in future.

\begin{acknowledgments}
The work is supported by the German Research Foundation (DFG) by grant Schm 2657/3. It is a pleasure to thank Michael Schmiedeberg, Liesbeth Janssen and Shiladitya Sengupta for useful discussions. The author also thanks Mohan Janakiraman for his inputs in the manuscript. 
\end{acknowledgments}


\begin{thebibliography}{100}

\bibitem{angell}
C. A.~Angell, {\it Science}, 1995, {\bf 267}, 1924-1935.

\bibitem{debenedetti}
P. G.~Debenedetti and F. H.~Stillinger, {\it Nature (London)}, 2001, {\bf 410}, 259--267.

\bibitem{richardetal}
P.~Richard, M.~Nicodemi, R.~Delannay, P.~Ribi{\'e}re and D.~Bideau, {\it Nature Materials}, 2005, {\bf 4}, 121-128.

\bibitem{hunterandweeks}
G. L.~Hunter and E. R.~Weeks, {\it Rep. Prog. Phys.}, 2012, {\bf 75}, 066501

\bibitem{puseyandmegen}
  P. N.~Pusey and W.~van Megen, {\it Nature}, 1986, {\bf 320}, 340--342.
  
\bibitem{parisiandzamponi}
G. Parisi and F. Zamponi, {\it Rev. Mod. Phys.}, 2010, {\bf 82}, 789-829.

\bibitem{weeksetal}
E. R. Weeks, J. C. Crocker, A. C. Levitt, A. Schofield, and D. A. Weitz, {\it Science}, 2000, {\bf 287}, 627--631.

\bibitem{goetze}
W. G{\"o}tze, {\it J. Phys.: Condens. Matter}, 1999, {\bf 11}, A1.

\bibitem{schweizer}
K. S. Schweizer, {\it J. Chem. Phys.}, 2007, {\bf 127}, 164506.

\bibitem{cohenandturnbull}
M. H. Cohen and D. Turnbull, {\it J. Chem. Phys.}, 1959, {\bf 31}, 1164-1169.

\bibitem{turnbullandcohen}
D. Turnbull and M. H. Cohen, {\it J. Chem. Phys.}, 1970, {\bf 52}, 3038-3041. 

\bibitem{cohenandgrest}
M. H. Cohen and G. S. Grest, {\it Phys. Rev. B}, 1979, {\bf 20}, 1077-1098.

\bibitem{weeksandweitz1}
E. R. Weeks and D. A. Weitz, {\it Phys. Rev. Lett.}, 2002, {\bf 89}, 095704.

\bibitem{weeksandweitz}
E. R. Weeks and D. A. Weitz, {\it Chem. Phys.}, 2002, {\bf 284}, 361-367.

\bibitem{doliwaandheuer}
B. Doliwa and A. Heuer, {\it Phys. Rev. Lett.}, 1998, {\bf 80}, 4915 - 4918.

\bibitem{ciamaraetal}
P.~Ciamara, R.~Pastore, and A.~Coniglio, {\it Sci. Rep.}, 2015, {\bf 5}, 11770.

\bibitem{berthierandbiroli}
L.~Berthier and G.~Biroli, {\it Rev. Mod. Phys.}, 2011, {\bf 83}, 587-645.

\bibitem{glotzer}
S. C. Glotzer, {\it J. Non-Cryst. Solids}, 2000, {\bf 274}, 342 - 355.

\bibitem{ediger}
M. D. Ediger, {\it Annu. Rev. Phys. Chem.}, 2000, {\bf 51}, 99 -128.

\bibitem{cammarotaetal}
C. Cammarota {\it et al.}, {\it Phys. Rev. Lett.}, 2010, {\bf 105}, 055703.

\bibitem{kobetal}
W. Kob {\it et al.}, {\it Phys. Rev. Lett.}, 1997, {\bf 79}, 2827-2830.

\bibitem{donatietal1}
C. Donati {\it et al.}, {\it Phys. Rev. Lett.}, 1998, {\bf 80}, 2338-2341.

\bibitem{donatietal}
C. Donati {\it et al.}, {\it Phys. Rev. E}, 1999, {\bf 60}, 3107-3119.

\bibitem{starretal}
F. W. Starr, J. F. Douglas, and S. Sastry, {\it J. Chem. Phys.}, 2013, {\bf 138}, 12A541. 

\bibitem{sastryetal1}
S. Sastry, D. S. Corti, P. G. Debenedetti, and F. H. Stillinger, {\it Phys. Rev. E}, 1997, {\bf 56}, 5524-5532.

\bibitem{sastryetal2}
S. Sastry, T. M. Truskett, P. G. Debenedetti, S. Torquato, and F. H. Stillinger, {\it Mol. Phys.}, 1998, {\bf 95}, 289-297.

\bibitem{maitietal}
M. Maiti, A. Laxminarayanan, and S. Sastry, {\it Eur. Phys. J. E}, 2013, {\bf 36}, 5.
 
\bibitem{rapaport} 
D. C. Rapaport, {\it J. Comp. Phys.},1980, {\bf 34}, 184-201. 

\bibitem{Brambillaetal}
G. Brambilla {\it et al}, {\it Phys. Rev. Lett.}, 2009, {\bf 102}, 085703.


\bibitem{maitiandsastry}
M. Maiti and S. Sastry, {\it J. Chem. Phys.}, 2014, {\bf 141}, 044510.

\bibitem{starretal1} 
F. W. Starr, S. Sastry, J. F. Douglas, and S. C. Glotzer, {\it Phys. Rev. Lett.}, 2002, {\bf 89}, 125501.

\bibitem{kumaretal}
S. K. Kumar, G. Szamel, and J. F. Douglas, {\it J. Chem. Phys.}, 2006, {\bf 124}, 214501.

\end{thebibliography}
\end{document}